\documentclass{article}

\usepackage{microtype}
\usepackage{graphicx}
\usepackage{subfigure}
\usepackage{booktabs} 
\usepackage{array}

\usepackage[hyphens]{url}
\usepackage[breaklinks]{hyperref}

\usepackage[preprint]{icml2026}

\usepackage{amsmath}
\usepackage{amssymb}
\usepackage{mathtools}
\usepackage{amsthm}

\usepackage[capitalize,noabbrev]{cleveref}

\newcolumntype{L}[1]{>{\raggedright\let\newline\\\arraybackslash\hspace{0pt}}m{#1}}
\newcolumntype{C}[1]{>{\centering\let\newline\\\arraybackslash\hspace{0pt}}m{#1}}
\newcolumntype{R}[1]{>{\raggedleft\let\newline\\\arraybackslash\hspace{0pt}}m{#1}}

\theoremstyle{plain}

\theoremstyle{definition}

\theoremstyle{remark}

\usepackage[textsize=tiny]{todonotes}

\usepackage{graphicx}
\graphicspath{ {./images/} }

\usepackage[compatibility=false]{caption}
\icmltitlerunning{Yes, But Not Always. Generative AI Needs Nuanced Opt-in.}

\begin{document}

\twocolumn[
\icmltitle{Yes, But Not Always. Generative AI Needs Nuanced Opt-in.}



\icmlsetsymbol{equal}{*}

\begin{icmlauthorlist}
\icmlauthor{Wiebke Hutiri}{yyy}
\icmlauthor{Morgan Scheuerman}{yyy}
\icmlauthor{Shruti Nagpal}{yyy}
\icmlauthor{Austin Hoag}{yyy}
\icmlauthor{Alice Xiang}{yyy}
\end{icmlauthorlist}

\icmlaffiliation{yyy}{Sony AI}

\icmlcorrespondingauthor{Wiebke Hutiri}{wiebke.hutiri@sony.com}

\icmlkeywords{Machine Learning, ICML}

\vskip 0.3in
]



\printAffiliationsAndNotice{}  

\begin{abstract}
This paper argues that a one-size-fits-all approach to specifying consent for the use of creative works in generative AI is insufficient. Real-world ownership and rights holder structures, the imitation of artistic styles and likeness, and the limitless contexts of use of AI outputs make the status quo of binary consent with opt-in by default untenable. To move beyond the current impasse, we consider levers of control in generative AI workflows at training, inference, and dissemination. Based on these insights, we position inference-time opt-in as an overlooked opportunity for nuanced consent verification. We conceptualize nuanced consent conditions for opt-in and propose an agent-based inference-time opt-in architecture to verify if user intent requests meet conditional consent granted by rights holders. In a case study for music, we demonstrate that nuanced opt-in at inference can account for established rights and re-establish a balance of power between rights holders and AI developers. 
\end{abstract}

\section{Introduction}
\label{sec:introduction}



Generative AI has experienced significant backlash due to the large-scale appropriation of digital assets as training data~\cite{jayachandran_traversing_2023}. Particularly in the creative industries, generative AI development practices pose an acute threat to the livelihoods of artists and to a thriving cultural ecosystem in which creative expression and production are incentivized. Thousands of individual artists and their representative organizing bodies across the world have publicly condemned the unauthorized use of their work to train AI~\cite{fairlytrained_statement_2024}, and have launched lawsuits to contest the unjust enrichment of AI developers at the cost of creatives~\cite{samuelson_generative_2023}. Central to creators' demands are issues of consent, opt-in, and control, which are a prerequisite for compensation~\cite{deladurantaye_control_2025}.

Protecting original creative work ensures that creators retain control and financial benefit from their work, thus incentivizing the ongoing creation and sharing of content~\cite{tabahriti_musicians_2025}. Copyright protects the interests of creators by granting them ownership of their Intellectual Property (IP) and control over how it is copied, distributed and adapted. Copyright law has been designed to offer a legal basis for creators' protection, balanced against public interest and fair use to allow others to build on creative work~\cite{hardinges_human_2025}. Generative AI disrupts this balance. Models are trained on vast quantities of data~\cite{gadre_datacomp_2023}, which are predominantly scraped from online sources~\cite{longpre_consent_2024}. This data acquisition approach has resulted in questions of consent and copyright being ignored by AI developers. Developers try to position that their claims to, and repurposing of, all data is fair use and in the public interest~\cite{sobel_artificial_2017}. As a natural consequence of generative AI's development process, public discourse and research on consent and generative AI has thus focused on copyright and opting out of AI training. Questions around the fair use doctrine in the US and regulations around text and data mining in the EU, UK, Singapore, and Japan have been central to these debates~\cite{sag_globalization_2024, deladurantaye_control_2025}.

Proposed legislative interventions to copyright challenges during training include licensing, meaningful compensation, opt-out, and transparency~\cite{senderdoff_generative_2025, noauthor_copyright_2025, deladurantaye_control_2025}. Litigations over copyright remain the major legal avenue that rights holders pursue to assert control over the ingestion of IP in AI training data. However, these interventions offer rights holders only limited levers to influence their training data concerns. Poor dataset transparency~\cite{scheuerman_treading_2026}, challenges around implementing opt-outs~\cite{newton-rex_insurmountable_2024}, and the technical difficulty of associating AI outputs with data samples in the training set through attribution~\cite{serra_supervised_2025} make it difficult for rights holders to assess the extent to which their IP has been ingested into models and to exert control over its removal. Moreover, as the viral Ghiblification trend early in 2025 highlighted~\cite{jon_ghiblification_2025}, generating outputs in the style of well-known artists or studios is a popular use case for generative AI. Yet, Creating secondary work \textit{in the style of} an original work is typically not considered a copyright infringement~\cite{samuelson_generative_2023}. 

With copyright protection being uncertain, creators are taking measures to protect their work into their own hands. Increasingly, creators and content owners are modifying their terms of service, restricting web crawlers, putting up paywalls, taking down content, and using tools aimed at model poisoning~\cite{weatherbed_how_2024}. Guidelines on \textit{how to stop your work from being used by AI} are already commonplace~\cite {winter_how_2024}. While these actions present a clear indication that creators demand control, many content owners do not trust that their preferences on whether to be included in AI training will be enacted properly~\cite{bui_optouts_2022}. These challenges are particularly pertinent for individual rights holders and institutions with limited resources to litigate, such as cultural heritage organizations~\cite{lehmann_position_2025}. Empirical studies show that creators' concerns about their content control preferences not being respected are justified, as preference statements are frequently unchecked, ignored or intentionally circumvented~\cite{cui_odyssey_2025, kim_scrapers_2025}. From the perspective of AI developers, creators' resistance to being scraped diminishes diverse and high-quality content available to train AI models~\cite{longpre_consent_2024}. The status quo is thus a lose-lose situation for creators and for the development of AI, which could support creative processes, production, and connect audiences to artists~\cite{miltner_possibilities_2024}. 

Current approaches to opt-in focus solely on training data, treat opt-in as a once-off, binary choice, place the burden to opt-out on rights holders, and assign control over adherence to opt-out to AI developers~\cite{newton-rex_insurmountable_2024}. To move beyond this impasse, alternative approaches to opt-in and control for creators are needed. The significance of user inputs is increasingly recognized in the AI ecosystem, with mechanisms like data flywheels~\cite{vincent_data_2025, gurkan_contracting_2022} and inference-time attribution~\cite{morreale_attributionbydesign_2025, kim_generation_2025} emerging to capture value in this stage of the AI life cycle. For language generation models, retrieval augmented generation (RAG)~\cite{lewis_retrievalaugmented_2021} is an inference-time intervention to augment user inputs with online search capabilities. Prior research has also argued that inference-time interventions are necessary for managing copyright risks~\cite{pan_position_2025}. Despite these trends, user inputs have thus far not been considered as a point of intervention for opt-in.

\textbf{This paper positions that opt-in should not be treated as a binary choice, but a context-specific, highly nuanced decision process vital for artistic autonomy, cultural sovereignty, and a thriving creative ecosystem. We advocate that generative AI needs mechanisms that enable nuanced consent with opt-out by default, and propose an inference-time solution to achieve this.}

\section{Opt-Out Means Opt-In by Default}
\label{sec:background}


AI developers have conveniently assumed \textit{opt-in by default}: that creative content accessible on the internet is freely available as a resource to be used in AI development~\cite{mackinnon_infrastructural_2025}. In response, rights holders and creators are turning to \emph{opt-out} protocols and metadata to restrict the use of their content in generative AI~\cite{dinzinger_survey_2024}. 

\begin{table*}[hbt]
    \centering
    \small
    \begin{tabular}{L{0.18\linewidth}L{0.4\linewidth}L{0.14\linewidth}L{0.18\linewidth}} \midrule
    \textbf{Proposal} & \textbf{Implemented via} & \textbf{Control Level} & \textbf{Explicit AI Preferences\newline (i.e. consent options)} \\ \midrule
    robots.txt (REP) & file at root & site & No \\
    Robots meta tags (REP) & HTML head & page & No \\
    X-Robots-Tag (REP) & HTTP header & content & No \\
    learners.txt & REP extension & site, content & train \\
    Content signals & REP extension & site & train, output \\
    AI Preferences & REP extension & site, page, content & train, output, search \\
    ai.txt & file at root & site, content & train, output*, search* \\
    TDMRep & file at root, HTML head, HTTP header, file metadata & site, page, content & custom \\
    ODRL & ontology & NA & train*, output*, search* \\
    CC signals & NA & content collection & NA \\
    \end{tabular}
    \caption{Comparison of AI-related content control proposals that operationalize AI preferences with diverse consent options. * denotes finer grained consent options that can be mapped to the captured higher level AI preferences}
    \label{tab:ai_content_control}
    \vspace{-1em}
\end{table*}

\subsection{Existing Opt-Out Mechanisms}

First proposed in 1994, the Robots Exclusion Protocol (REP)~\cite{koster_robots_2022} defines how automatic crawlers may access content on the internet. It is implemented via a robots.txt file in the root directory of a website. The file allows a webmaster to limit the sub-domains that bots can access. Additionally, robots meta tags can be used in HTML to specify how crawlers can use, index, archive and display content in search results. An X-Robots-Tag can be used for non-HTML files, like audio and images. The REP and instructions in robots.txt were initially intended as a guideline and not a legally binding contract, but REP has been cited in copyright disputes~\cite{noauthor_web_2007}, taking on a new, legal role \cite{yang_using_2010}. 

The REP has limitations in its role as a legal contract, especially with regards to copyright~\cite{yang_using_2010, nico_generative_2025}. The directives that the REP provides are insufficient for covering all associated rights that can be infringed on. Moreover, a webmaster's intent behind directives and tags can be ambiguous, page redirects can eliminate previously stated preferences, and rights holders do not always have the ability to influence the robots.txt file. Several new proposals for data use preferences have been tabled to overcome these shortcomings (see \cref{tab:ai_content_control}). For example, Cloudflare's Content-Signal directive~\cite{cloudflare_content_2025}, the learners.txt file~\cite{ippolito_donottrain_2023} and IETF's draft AI Preferences standard~\cite{illyes_associating_2025} propose AI-specific extensions to the REP. TDMRep~\cite{lemeur_tdm_2024}, a W3C web protocol to declare preferences in a JSON file stored in a site's \texttt{/.well-known} directory, enables rights holders to specify nuanced rules for rights reservations and usage policies. Usage policies can be specified in a standardized manner with the Open Digital Rights Language (ODRL)~\cite{mcroberts_open_2014}, an endorsed W3C recommendation which was explicitly created for digital rights management. \citet{li_aitxt_2025}'s ai.txt file also offers a domain-specific language that captures preferences for a range of AI actions, including AI training, AI outputs, and actions agents may take, such as clip, describe, rephrase, transcribe, or translate.

\subsection{Shortfalls of Opt-Out}

Webmasters increasingly rely on the REP to restrict access to online content out of fear of content being scraped for AI training~\cite{longpre_consent_2024, dinzinger_longitudinal_2024}, especially in news and finance domains~\cite{cui_odyssey_2025}. However, studies have found cases where REP directives were ignored or violated, and where web content was memorized~\cite{cui_odyssey_2025}. For example, researchers have found hundreds of millions of materials with copyright or do-not-scrape notices in CommonPool~\cite{gadre_datacomp_2023}, a large-scale dataset for vision-language tasks~\cite{lee_how_2025}. Empirical data suggests that adherence decreases as directives become more restrictive, that many bots check robots.txt infrequently, and that malicious bots attempt to circumvent robots.txt by posing as bots with fewer access restrictions~\cite{kim_scrapers_2025}. If opt-outs are not respected, rights holders are coerced to opt-in by default. 

Beyond being willingly ignored by developers, opt-out of training data also has practical challenges that render it ineffective in many situations~\cite{newton-rex_insurmountable_2024}. Opt-out consumes massive administrative overhead, as copyrighted works are ubiquitous on the internet. It is unachievable to opt-out all instances where a copyrighted work appears. Similarly, it is not possible for rights holders to opt-out of every AI system. Even if it was possible, the ever-changing landscape of technology makes maintaining opt-outs burdensome, in particular for individual creators. Prior research has recommended that collective or representative consent mechanisms could support creative professionals with managing opt-outs~\cite{kyi_governance_2025}. Yet, as long as AI developers retain control over the implementation of opt-out mechanisms, they will remain difficult to enforce and only effective if accompanied by transparency requirements. 

Conceptually, opt-out does not reflect the opt-in nature of copyright~\cite{kim_why_2025}. Once copyrighted works have been ingested as training data into models, opt-out is inffective. The problem compounds in second and third generation models trained on synthetic data, where the original rights holders whose work made the synthetic data possible are no longer traceable. Training-time opt-out is thus a binary choice between consenting to all future uses of models in perpetuity, and missing out on potential beneficial uses of generative AI entirely. This binary choice leaves no room for negotiating terms and neglects to account for complex rights scenarios that are typical of multimodal data. 

\section{In Support of Nuanced Opt-In}
\label{sec:framework}

In this section we support our call for nuanced opt-in by looking at the stakeholders involved, the datafication of style and likeness, and the contexts of use for AI outputs.

\subsection{One Work, Many Rights Holders} 

Rights holders are individuals or organizations who own the IP of a work and legally control how it can be used~\cite{rahmatian_copyright_2011}. When a rights holder opts in, they consent to the use of their creative work. For example, an individual illustrator who holds the rights to all their illustrations can choose whether or not to opt-in their works to an AI system. However, rights can quickly become complex. For example, a single music track might involve multiple rights holders: multiple singers, a songwriter, a producer, sound engineers, even samples from another musical artist. It is common for art forms, particularly commercialized ones, to have highly complex rights~\cite{kostova_creators_2017}. Moreover, IP ownership and associated rights are not static. They can lapse, be sold or transferred, for example after the death of the original rights holder~\cite{cianisciolla_no_2025}. 

There are limits to relying on IP protection alone when considering opt-in for generative AI. Copyright differs across countries and the sharing of works online makes enforcing country-specific laws legally complex~\cite{eisenberg_artists_2025, fitzgerald_country_2015}. Some creators may not even look to copyright for protection. In the high-end art market it has been argued, for example, that the value of authenticity outweighs the threat of copies~\cite{adler_why_2018}. Personal, cultural, spiritual and social incentives to create art can also outweigh commercial interests. Some creative works, like, folk songs, traditional music and struggle songs often times cannot be traced back to individual creators. The communities that cherish them may not view them as individual property~\cite{bachner_facing_2005}. The collective custodianship of communities over their cultural goods can thus stand in opposition to the assignment of exclusive ownership rights~\cite{bruhn_identifying_2014}. 

Questions around multi-party ownership and collective custodianship extend to the context of AI~\cite{kirui_epistemics_2025}. Opt-in must be nuanced to offer conditional consent that reflects the interests of diverse stakeholder groups, so that an individual painter who posts their works to social media as a hobby can choose different consent conditions to a signed musician with numerous collaborations.  

\subsection{The Datafication of Style and Likeness} 
\label{datafication}

Like many other aspects of human life, creative works have been digitized and datafied so that they can be processed and automated~\cite{mayer-schonberger_big_2014, mejias_datafication_2019}. Generative AI has been particularly adept at expanding processing opportunities from clearly defined aspects of creative work, like colours, pitch, or vocabulary, to aspects that are high-dimensional, non-linear and difficult to define exactly. At the root of many artists' concerns with generative AI lie two such aspects of creative work: namely the non-consensual use and monetization of their \emph{likeness} and \emph{signature style}~\cite{benton_artists_2025}. Signature style is the distinctive and recognizable way an individual artist expresses themselves in their work~\cite{genova_significance_1979}. It may take years for an artist to develop a signature style, yet style remains difficult to define and quantify. While users like to prompt for works in the style of an artist, many artists do not want their signature style to be used in generative AI outputs \cite{sobel_elements_2024}. Historically it has not been viewed as a copyright infringement for a derivative work to be based on the style of an original, provided that the secondary work does not appropriate significant or expressive portions of the original~\cite{samuelson_generative_2023}. 

Similarly, many creatives have concerns about how their likeness is used in generative AI outputs, irrespective of whether they are the rights holder. Deepfakes in particular are an area of concern. In the US, publicity rights prevent the unauthorized use of people's likeness for commercial gain. In other countries, personality rights protect `one's name and own image'~\cite{abdullahahmed_critical_2024}. However, these rights have limitations when it comes to generative AI. Voice actors, for example, have resisted the limitless use of their voices in generative AI, as it directly impacts their ability to generate income~\cite{hutiri_not_2024}. With social media enabling rapid, global distribution of AI generated content, mainstream artists like Jorja Smith have encountered viral deepfakes being passed off as legitimate \cite{savage_jorja_2025}. 

Due to the new capabilities of generative AI, the datafication of style and likeness have become prevalent. Creators want to have control over them, but they do not fall under the purview of copyright. Copyright is thus unlikely to protect signature styles. Currently only Denmark has changed their copyright law to protect people from their likeness being used in generative deepfakes~\cite{bryant_denmark_2025}. Nuanced opt-in is an opportunity for creators to retain control over their signature style and likeness.

\subsection{Limitless Contexts of Use}

Generative AI has an infinite repertoire of outputs that can be created. With social media, generated outputs can reach a near global audience. This makes the potential contexts of use of generative outputs near limitless. Yet, rights holders and creators may not want their work to be used by anyone, for any purpose, be it profane, mundane, insane or inhumane. Creators already have a history of setting terms of use for their work through licensing agreements. Creative Commons licenses, for example, attend to questions of commercial and non-commercial use. Other purposes might also be of interest, like educational or research uses \cite{kapitzke_copyrights_2011,kiel-chisholm_rise_2006}. Meanwhile, regional laws can protect certain uses, regardless of the rights holders' preferences. For example, in the United States, ``services at a place of worship or other religious assembly'' can use literary or musical works without copyright infringement \cite{delaurenti_copyright_2024}.

Rights holders of AI inputs need to be able to limit how and where the outputs of generative AI can be used. Creatives may wish to enable or disable specific uses or sharing practices, especially if those uses might violate their personal values, damage their brand, or directly compete with their livelihoods. Providing consent only if particular distribution practices are adhered to could thus implicate specific platforms---like YouTube, Spotify, Instagram, and TikTok---to abide by artist terms. If creators cannot restrict the context of use of their work, they loose autonomy and self-determination. Opt-in to generative AI should enable these types of controls, so that rights holders can limit how generated outputs that consume their work are used.

\section{How to Enable Nuanced Opt-in}
\label{sec:operationalising}

Next, we describe how to enable nuanced opt-in that accounts for diverse creative stakeholders, grants control over style and likeness, and that is sensitive to contexts of use. 

\subsection{Expanding Rights Holders' Locus of Control}

\begin{figure}[bt]
    \centering
     \includegraphics[width=\linewidth]{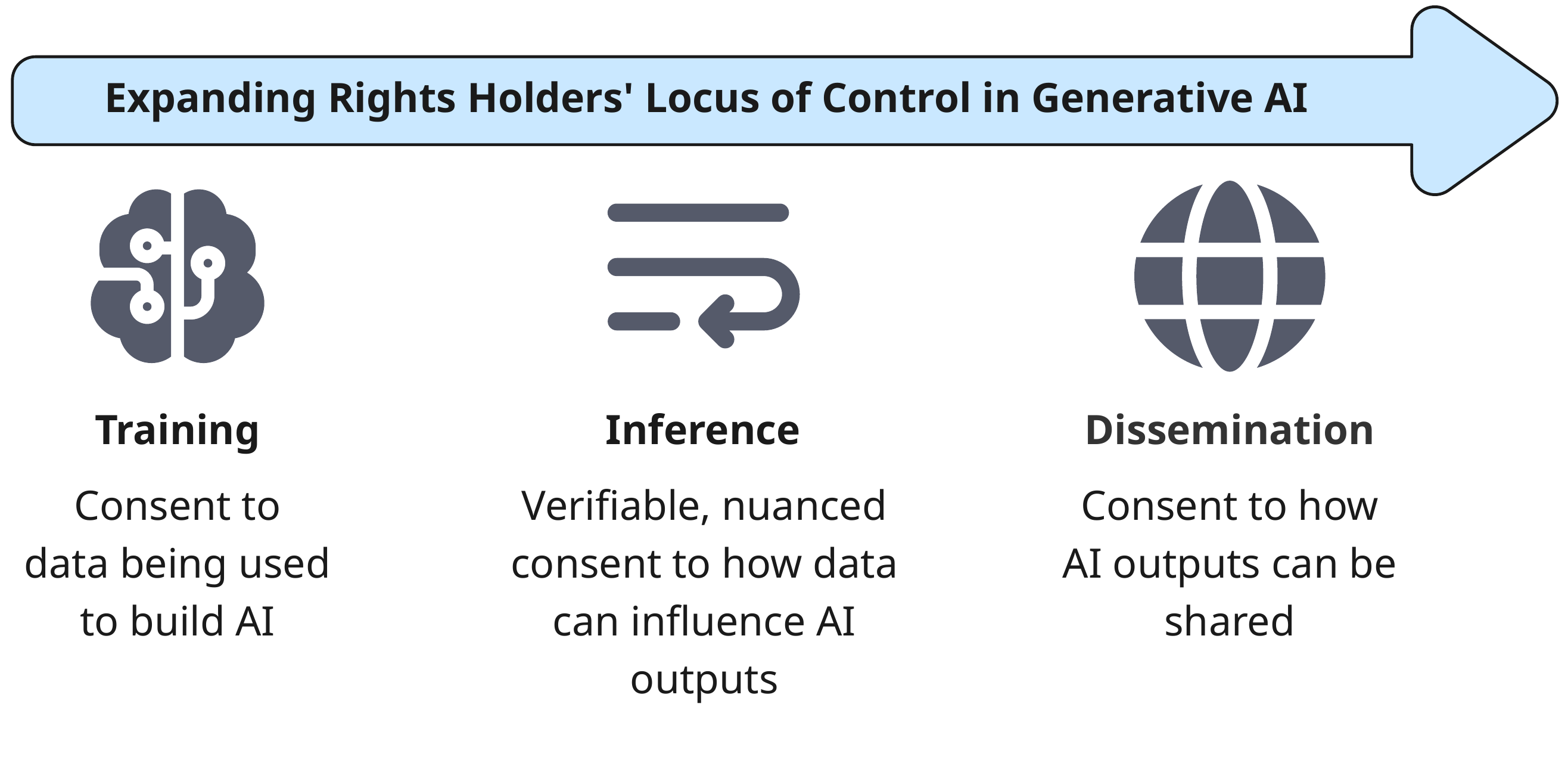}
    \caption{Expanding rights holders' locus of control from training data to inference and dissemination}
    \label{fig:expanding_control}
\end{figure}

AI training is only one leverage point for rights holders to control how their work is used in and by AI. By taking a life cycle view on generative AI, we identify additional opt-in levers for rights holders. Figure~\ref{fig:expanding_control} visually demonstrates how consent can be leveraged at different phases in the AI life cycle -- during training, inference and dissemination -- to exercise control over AI. As data plays different roles in each phase of the lifecycle, this also leads to different implications to what rights holders consent to and how this can be steered through opt-in.
 
At training time, IP is directly ingested into AI models. Consequently, for a rights holder to opt-in to AI training, means that they give consent to their IP being used to build AI. A disadvantage of valuing data for training-time use, is that there is no direct correlation between the value that a creative work has to humans and the value that the corresponding data sample has to a model. Offering high value IP, such as music back catalogues, for training thus has limited upside for rights holders, but large potential down sides. Once the data distribution has been learned, subsequent models can be trained on synthetic data produced with the initial model, rendering the original work irrelevant. Ultimately, training-time opt-in is a binary choice that offers rights holders no direct mechanisms to control how, where and for how long models are used, or what happens with model outputs. Opt-in to AI training thus requires a disproportionate amount of trust from rights holders that developers will honour agreed upon terms indefinitely. While opting in to AI training remains an important lever of control for rights holders to affirm their willingness to have their IP used to build AI, additional nuance and control is needed to safeguard the creative ecosystem, the economic value of creative work and the autonomy of creators.

At inference, a trained model is called with prompts via user inputs. Inference thus presents an additional point of leverage where rights holders can exert control over how their work is used to generate outputs. Instead of using guardrails to block user inputs that reference protected entities, verifying opt-in and consent during inference can unlock the use of IP in permitted scenarios, giving rights holders fine-grained control to specify consent conditions. Inference-time solutions make it possible to assume opt-out by default, while requiring explicit opt-in for any use of creative content in the AI lifecycle. Following opt-in verification during inference, the final opt-in step is to establish how a user intends to disseminate the generated output. Further restrictions can be placed on how AI outputs are allowed to be shared, given the user's dissemination intent. 

By expanding rights holders' locus of control from training to inference and dissemination, multiple sources of compensation are unlocked: upfront payment for training data licensing, recurring compensation for IP use via user inputs and shared revenue through distribution of generated outputs. This positions nuanced opt-in as a critical component of generative AI compensation mechanisms, that specifies and enables a value exchange between rights holders and users while giving rights holders control over their IP.



\subsection{Conceptualizing Nuance for Opt-in} 

We now consider what \textit{nuance} means for opt-in of creative work. Our assumption is that users generate outputs they want, rather than arbitrary content. User inputs can thus be viewed as an expression of user intent~\cite{choi_understanding_2025}. Conventionally, inputs are natural language text prompts, but can also be multimodal. We further assume that a rights holder opts in to a particular user intent if and only if all conditions under which they are willing to grant consent are met. For nuanced opt-in to work, it must be possible to obtain consent conditions from rights holders in a realistic time-frame. Consent conditions should also capture some underlying value that is pertinent to the IP or rights holders. 

\begin{table}[bt]
    \centering
    \small
    \setlength{\tabcolsep}{4pt}
    \setlength\extrarowheight{3pt}
    \begin{tabular}{llL{0.5\linewidth}} \toprule
         & Generic & Reference a general category, \textit{e.g. anime, ceremonial} \\ 
         \textbf{Descriptors} & Specific & Reference a unique entity, \textit{e.g. Studio Ghibli, ``Pomp and Circumstance"} \\ 
         & Original & New creative input, \textit{e.g. user photo, user song} \\ \midrule
         & Generic & Modify descriptor to resemble a general category \\
         \textbf{Transform's} & Specific & Modify descriptor to resemble a unique reference \\ 
         & Foundation & Modify descriptor based on rules or theory, \textit{e.g. increase blue hues, b-flat minor} \\ \midrule
         & Quality & Value-determining output quality, \textit{e.g. high res, 10sec}\\ 
         \textbf{Qualifiers} & Distribution & Target audience and platforms, \textit{e.g. Instagram}\\ 
         & Purpose & Output ownership and use, \textit{e.g. social sharing}\\ \bottomrule
    \end{tabular}
    \caption{Abstract prompt syntax to categorize users' intent and conceptualize nuanced consent conditions for opt-in.}
    \label{tab:opt_in_taxonomy}
\end{table}

We first define an abstract prompt syntax to categorize the core components available to users to express their generation intent. At the top level, intent components can be categorized as descriptors, transformations and qualifiers, as shown in Table~\ref{tab:opt_in_taxonomy}. \textbf{Descriptors} present the primary user intent and provide the minimal attributes to condition the output. Descriptors' attributes can be modified with \textbf{transformations}. Lastly, \textbf{qualifiers} specify attributes of the output that are not directly connected to the input, but that may modify the output's value or usability. We illustrate the syntax with the prompt \textit{``Turn this image into a Ghibli anime style}". \textit{``This image"}, i.e. an image provided by the user, is the descriptor, to which the user wants to apply a style transformation. Possible options of qualifiers that a user could have specified in our example would be \textit{``in high resolution"} or \textit{``for an Instagram post"}.

On the next level, we distinguish between generic and specific descriptors and transformations. \textbf{Generic} descriptors and transformations reference a general category, such as an artistic style or era (e.g. anime). \textbf{Specific} descriptors and transformations, on the other hand, reference unique entities, in particular creative works (e.g. albums, songs) or people. Descriptors can also be \textbf{original} if they are a new creative input, such as a photo or song recorded by the user. We further describe transformations as \textbf{foundational} if they are based on rules or theoretical knowledge that the user specifies. For example, a user may prompt for an increase in blue hues in an image. Such a transformation, while specific, does not require a reference to a unique entity. Qualifiers can specify the desired output \textbf{quality} (e.g. image resolution or song length), \textbf{distribution} to target audiences and platforms (e.g. Instagram) and the \textbf{purpose} of use of the output (e.g. social sharing). Qualifiers can alter the commercial value of the output and thus impact compensation.

Specific descriptors and transformations reference unique entities, which makes it possible to obtain consent conditions and consequently perform opt-in checks and verification for their use. \textit{With a nuanced approach to opt-in, rights holders can thus decide whether they are willing to be referenced as specific descriptors, whether they consent to be referenced to transform descriptors or to have their own work modified or combined with other works}. Moreover, rights holders can provide fine-grained limitations on which of their work can be used as reference, and even narrow this further to include or exclude specific transformations to specific parts of a work. Qualifiers can be used to specify additional consent conditions that either permit limited use or prescribe how work can be used to generate outputs. In contrast to specific user intent, generic descriptors and transformations are inherently abstract and do not refer to entities with unique rights. Through feedback and recommendations, users can be guided to reframe generic requests into specific descriptors or transformations. For original descriptors, we consider opt-in to be implicit. 



\subsection{Realizing Nuanced Opt-in at Inference}
\label{inference}

\begin{figure*}[tbh]
    \centering
    \includegraphics[width=\linewidth]{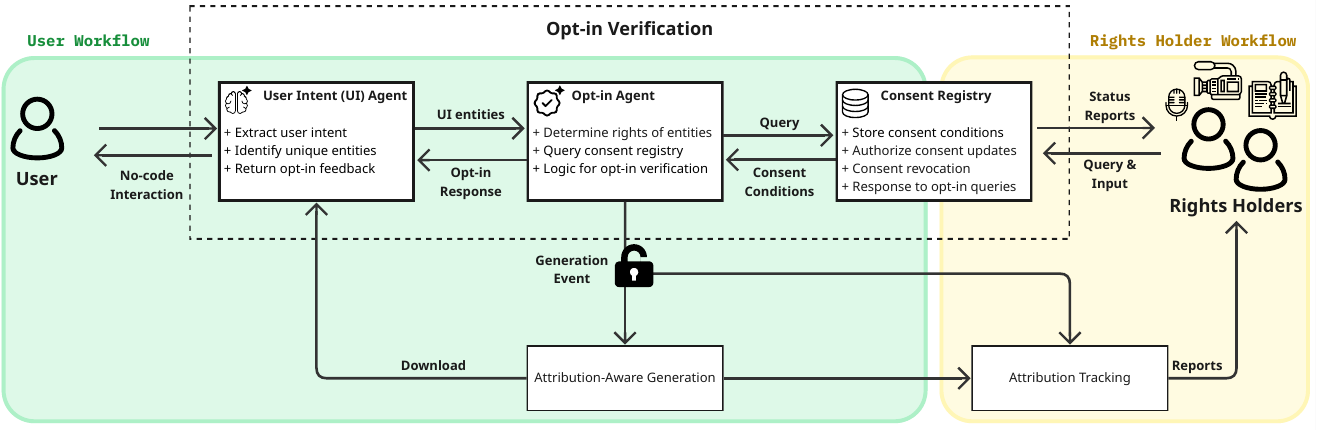}
    \caption{Agentic framework for inference-time opt-in verification to allow rights holders and creators to control how their work can be used to condition generative AI outputs.}
    \label{fig:inference_time_opt_in}
    \vspace{-1em}
\end{figure*}

By expanding the focus from training data to additional leverage points in the AI lifecycle, it is possible to shift from a binary opt-in by default paradigm towards nuanced opt-in at inference. Based on this insight, we propose the inference-time opt-in architecture in Figure~\ref{fig:inference_time_opt_in} to systematically verify unique entities and rights holders referenced in user prompts. Our proposed opt-in architecture respects \textit{conditional consent and allows rights holders and creators to control how their work, likeness and style can be used to condition generative AI outputs}. The architecture leverages AI agents and consists of three main components: i) a User Intent Agent, ii) an Opt-in Agent and iii) a Consent Registry.

\paragraph{User Workflow.} The generation process is initiated by a user who interacts with a no-code input interface to express their generation intent. The User Intent (UI) Agent manages this interaction by extracting the user's intent from the input, identifying unique entities referenced in the input and passing them on to the Opt-in Agent. In the return loop, the UI Agent also provides opt-in feedback to the user. The Opt-in Agent determines if the extracted entities have associated rights and prepares a query that it passes to the Consent Registry. Based on the consent conditions returned from the registry, the Opt-in Agent verifies if the user's intent satisfy rights holders' conditions for use, and passes the outcomes of this verification back to the UI Agent. If the Opt-in Agent confirms opt-in, a generation event is unlocked and attribution-aware generation can take place. Attribution of generation events is tracked, such that the results of the opt-in verification process and any subsequent downloads or distributions of generated outputs are logged and traceable. The Consent Registry is a database that stores consent conditions of rights holders, authorizes consent modifications and revocations. It also provides responses to queries from the Opt-in Agent or rights holders. 

\paragraph{Rights Holder Workflow.} The opt-in verification architecture is based on four principles that elevate the autonomy of rights holders: 1) control 2) nuance, 3) time-boundedness and 4) transparency. The Consent Registry is designed with an opt-out by default as standard. That means that any rights holder or creative work that is not in the registry cannot be referenced and used to condition the generation process. Rights holders that wish to release some of their work for AI generation, can input their consent conditions to a fine degree of nuance into the Consent Registry. Consent is not treated as a permanent permission. Rights holders can modify or revoke consent, or grant it for a limited period of time. Furthermore, they can query the registry to obtain status updates on their consent conditions, and retrieve reports on queries to the registry by the Opt-in Agent. If opt-in is granted to a user request and a generation event takes place, the rights holder is attributed and eligible for compensation.










\section{Case Study: Nuanced Opt-In for Music}
\label{sec:case_study}


To illustrate how nuanced opt-in at inference can work, we present a case study for the music industry. 

\subsection{The Case for Nuance}
Music is typically multi-modal, combining audio-visual performances, sound recordings and different forms of written text, such as sheet music, lyrics and descriptions. More often than not, music involves multiple creators, from vocalists and instrumentalists to composers, sound engineers and producers. The same song can be played as a live performance or recorded. Recordings can be re-recorded at multiple points in time, with different performers and even major modifications such as a new melody for well-known lyrics. Recorded musical works can be original productions, but can also reuse and incorporate parts of previous recordings through sampling. The music industry commercializes the music of creators, with the best known creator groups being performers and songwriters. In addition to creators, rights holders in the music industry include labels, which fund, market and distribute music production, and publishers, which manage, protect and monetize lyrics and melodies~\cite{tschmuck_record_2016}. 

A single song can be decomposed into many elements: pitch, rhythm, duration, structure, dynamics, pulse, texture, tempo, timbre, tonality, and harmony. It can contain vocals and associated lyrics; it may have multiple instruments recorded as overlaying tracks or stems, containing elements like electric guitar, bass guitar, snare drums, keyboard, lead vocals, background vocals, and so on. Songs are sometimes broken up into sections, like a pre-chorus and a chorus. The sheer number of potential rights holders and musical elements in just one song highlights that opt-in goes beyond the consent conditions of single artists and calls for nuance.  

\subsection{From Rights to References and Consent}
It is quite clear that music rights holders are not a singular entity and that musical works comprise layers of creative work. Despite this complexity, creativity and innovation are hallmarks of music creation, which is supported by mature and established processes for assigning and managing rights of creative works. Music rights currently fall into two broad categories. Composition rights are granted to the fixed expression of musical works in a tangible form, such as lyrics and melodies~\cite{noauthor_copyright_2023}.They can in certain circumstances also apply to unique rhythms, harmonies and arrangements. Distinct from this are master and recording rights, which are granted to the sound recording that captures the audio of a particular performance in a fixed format~\cite{noauthor_copyright_2021}. Master rights typically pertain to the performance and to the capture and post-processing of the final recording.


When prompting, the user's intent may be to invoke a variety of specific references, including requests to imitate an artist's style or voice, an album's style or a song. For a song, a user may wish to refer to specific aspects of the song, such as the lyrics, melody, beat or individual instrumentals. Some specific references can be mapped directly to aspects of music that are protected by copyright. However, even for specific references that are not protected by copyright, rights holders can specify nuanced consent conditions that can be verified during inference. If conditions are met, opt-in is granted and the user request can lead to a generation event.


\subsection{``Sing this Song with Grimes's Voice"}

We close off with an example of how a music generation request would pass through our proposed opt-in verification architecture. The example is based on the consent conditions set out in a pilot partnership between artist Grimes, CreateSafe and TuneCore\footnote{\url{https://support.tunecore.com/hc/en-us/articles/16428915033492-Distributing-collaborations-with-GrimesAI}}. 

Consider a user who submits the following prompt ``\textit{Create a song from `Rolling in the Deep' with Grimes's Voice}". The user input contains a specific descriptor, `Rolling in the Deep', and a specific transformation, the reference to Grimes's voice. The User Intent Agent extracts these two unique entities and identifies the user's intent as creating a song from the entities. It passes this information on to the Opt-in Agent. The Opt-in Agent identifies `Rolling in the Deep' as a song and Grimes as artist. It then submits a query to the Consent Registry to retrieve the consent conditions for the two entities. `Rolling in the Deep' is not found in the registry, and a null result is returned for this request. Grimes, on the other hand, has entries in the Consent Registry. For Grimes, the following consent conditions are returned to the Opt-in Agent: specific descriptor - voice: permitted, combination with original descriptors: permitted. 

The Opt-in Agent now verifies if the user's request meets the consent conditions that have been retrieved. For Grimes, verification fails, as no permissions are granted for transformations. Thus, even though the user would be permitted to use Grimes's voice if they uploaded their own original sound recording and lyrics, opt-in is not granted for using Grimes's voice with another copyrighted work. There is no further need to check the conditions of `Rolling in the Deep', though opt-out by default implies that opt-in verification would have failed for this check. The Opt-in Agent returns the opt-in response to the User Intent Agent, which provides the feedback to the user and guides them to adjust their request to a permissible query. 


\section{Alternative Views}
\label{sec:discussion}


We raise three alternative views to our position on nuanced opt-in: implementation, bias and refusal. Nuanced opt-in can be difficult to implement due to practical and technical limitations. Open questions remain on how to enact nuanced opt-in when multiple and diverse creators and rights holders are involved. Instances of consent conflicts will need to be resolved, and aligning on approaches for doing this may be difficult~\cite{walquist_collective_2025}. Similarly, consent cannot be obtained directly for entire musical genres or artistic eras, making nuanced opt-in infeasible for generic generation requests. While generic requests can be turned into an artist recommendation or discovery process, it is not clear how best to do that. Even though large-scale opt-in databases for rights holders are seen as promising, they require industry buy-in to succeed~\cite{epstein_artificial_2025}. Governance and control challenges may arise in determining who should own and operate such a database, how the database might work across jurisdictions with different legal frameworks, and how to ensure interoperability~\cite{posth_federated_}. Technically, the extent to which accurate opt-in verification and policy adherence can be implemented has not been established. Gaming and adversarial attacks which abuse and violate opt-in specifications can pose further challenges~\cite{kim_automatic_2024, wang_stronger_2024}.

Bias remains a systemic concern for generative AI that cannot be solved with nuanced opt-in. User intent and interaction design will determine which works are referenced and thus which creators benefit. Inference-time opt-in in particular may lead to availability and popularity bias, with successful and well-known artists gaining, while independent, less popular, less known, or historically marginalized creators struggle to reap rewards \cite{kelly_pressure_2024}. Current model capabilities are also likely to pose limitations, as music processing and generation are focused on the Global North~\cite{mehta_missing_2024, ma_foundation_2024}. Ensuring that outputs are of high quality for input references across artists' genders, regions, languages, genres, styles, and popularity is important to ensure nuanced opt-in actually benefits all artists, not only those who are already established.

Many creatives are opposed to the use of AI in the creative sector, irrespective of the level of granularity at which consent can be granted~\cite{nath_we_2025,herington_musicians_2025,zhang_rise_2025}. Even if granted control over how their works can be used, creators may still see their livelihoods endangered if AI shifts markets away from original creation~\cite{kawakami_impact_2024}. AI has also been critiqued for its potential to homogenize creativity, given that it can only reference existing works \cite{schoenberg_how_2024,daryani_homogenizing_2025,anderson_homogenization_2024}. The proliferation of what critics label ``AI slop'' is already shifting user experiences on music streaming and social media platforms where many creatives have traditionally shared their works \cite{pendergrass_strategic_2025,madsen_when_2025}. Nuanced opt-in does not replace the need for more robust regulations on AI uses and improved labor protections for the creative sector~\cite{lee_rethinking_2022}. Nonetheless, in the face of ongoing development of generative AI, nuanced opt-in presents a substantial improvement to current opt-out approaches.

\section{Conclusion}
\label{sec:conclusion}

This paper positions the need for nuanced opt-in for generative AI. In support of this position, we highlight that creative works often times have many rights holders, that aspects of creative works like signature style and likeness extend beyond copyright and intellectual property, and that generative AI can be used for infinite purposes which rights holders may not endorse. Current proposals for opt-out lack the ability to protect creators and rights holders. This paper calls for alternatives that offer rights holders explicit and granular control mechanisms with conditional consent and opt-out by default. We propose nuanced, inference-time opt-in as a solution and present an agent architecture to realize this. We illustrate in a case study how nuanced opt-in at inference can centre the interests of creators in the music industry and re-establish a balance of power between rights holders and AI developers.

\bibliography{references, manual_references}
\bibliographystyle{icml2026}



\end{document}